\documentclass[twocolumn,showpacs,preprintnumbers,amsmath,amssymb]{revtex4}

\usepackage{graphicx}
\usepackage{dcolumn}
\usepackage{bm}

\begin{document}
\title
{Interference Effects on Kondo-Assisted Transport \\
through Double Quantum Dots
}

\author{
Yoichi Tanaka and  Norio  Kawakami
}

\affiliation{
Department of Applied Physics, Osaka University, Suita, Osaka 565-0871, Japan
}%

\date{\today}

\begin{abstract}
We systematically investigate electron transport through
double quantum dots with particular emphasis on
interference induced via multiple paths of electron propagation. 
By means of the slave-boson mean-field
approximation, we calculate the conductance, the local density of states,
the transmission probability in the Kondo regime at zero temperature.
It is clarified how  the Kondo-assisted transport
changes its properties when the system is 
continuously changed among the serial, parallel and  T-shaped double dots.
The obtained results for the conductance are explained in terms of the 
 Kondo resonances influenced by interference effects. 
We also discuss the impacts
due to the spin-polarization of ferromagnetic leads.
\end{abstract}


\maketitle

\section{Introduction}
Recent intensive investigations of electron transport through
quantum dot (QD) systems have uncovered 
novel correlation effects in nanoscale systems. 
In particular,
the observation
\cite{Gold,Cronen}
 of the Kondo effect in  QD
systems
\cite{Glaz,Ng,MW,Kawa,Ogu}
opened a new path for the investigation of
strongly correlated electrons, which has stimulated
further experimental and theoretical studies
in this field
\cite{Kouwen}.
More recently, a variety of QD systems, such as
 an Aharonov-Bohm ring with QD, double quantum dot (DQD) 
systems, etc,  have been fabricated.
An interesting new feature in these systems
is the interference effects induced via multiple paths of electron
propagation
\cite{Hack}. 
For example, such interference effects have been 
clearly observed in the Aharonov-Bohm ring with QD
\cite{Yaco,Buks,WG,Ji,Koba}.

When electron correlations due to the Kondo effect 
are affected by  such interference,
 transport properties exhibit remarkable properties.
There have been a number of theoretical works on DQD systems
in this context. For example, it was pointed out that the
DQD system shows the dramatic suppression of the Kondo-assisted 
transport due to the interference when the DQD is
arranged in the parallel geometry
\cite{Shimizu,Rosa,Boese,Imamura,Chen,LuYu,Lei,tanaka2}, 
whereas such interference
effects do not appear in the serial DQD
\cite{Lang,George,Rosa,Eto,Eto2,Busser1,Izumi,sakano,Jeong}.  
More recently, slightly different DQD systems
with the parallel geometry, where the two dots are connected via
the exchange coupling
\cite{LuYu} or the tunneling
\cite{Lei,tanaka2}, have been studied. It has been pointed out that the 
Kondo-assisted transport in these cases exhibits different properties
from the above simple DQD case without the inter-dot
coupling. The T-shaped DQD
\cite{Tae,Kikoin,Taka,Busser2,Apel,Corna} is 
another prototype of such correlated systems, for which the special
arrangement of the DQD provides an additional path of electron propagation, 
giving rise to the interference effects. This system also shows
somewhat different transport properties from the parallel
DQD case, but the detailed analysis has not been done systematically.
In any case, the interplay of the Kondo effect and the interference
provides intriguing phenomena due to electron correlations
in the DQD systems. It is thus interesting to systematically study how 
the interference together with the Kondo effect affects
characteristic transport properties in a variety of DQD systems.

Motivated by the above hot topics, we investigate Kondo-assisted 
transport properties in DQD systems,
 when the geometry of the system is systematically  changed among
the serial,
parallel 
and T-shaped DQD.
We pay our particular attention to the Kondo effect 
 under the influence of the interference due to different paths.
We also discuss the DQD systems with ferromagnetic (FM) leads.
This is stimulated  by  the recent extensive study of
spin-dependent transport through a QD coupled to
FM leads in the context of spintronics
\cite{Wang,Zhang,Sergu,Lopez,Koing,Utumi,Marti,JiMa,
BinDon,Choi,tanaka1,Ralph}.
Since the Kondo effect is sensitive to the
internal spin degrees of freedom, 
notable phenomena caused by the FM leads are expected
to appear in transport properties.

This paper is organized as follows. In the next section, we introduce
the model and briefly summarize the formulation based on 
the Keldysh Green function method. 
Then in Sec. \ref{sec:result}, we calculate the local
density of states (DOS) and the linear conductance
at zero temperature by the slave-boson 
mean field approximation
\cite{AC,Cole} for  the
serial, parallel and T-shaped DQD systems.
We discuss characteristic transport properties
induced by the Kondo effect 
under the influence of interference.
The effect of the spin-polarization due to the FM leads is
also addressed. A brief summary is given in the
last section.

\section{Model and Method}

\begin{figure}[h]
\begin{center}
\includegraphics[scale=0.35]{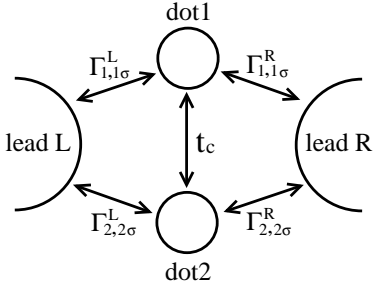}
\end{center}
\caption{ DQD system connected by 
 the inter-dot tunneling $t_c$:
$\Gamma^{\alpha}_{m,m\sigma} (\alpha =L,R$ and $m=1,2)$
represents the resonance width due to transfer
 between the $m$-th dot and the $\alpha$-th lead
for an electron with spin $\sigma $. By changing the 
ratio of tunneling amplitudes, we continuously
change the system among the serial, parallel and 
T-shaped DQD.
}
\label{modelP}
\end{figure}

We consider a DQD system shown schematically in Fig. \ref{modelP}
\cite{Shimizu,Rosa,Boese,Imamura,Chen,LuYu,Lei,tanaka2,Gueva}.
In the following discussions, the intra-dot  Coulomb
interaction is assumed to be sufficiently large, so that
double occupancy on each QD is
forbidden. This assumption allows us to use  a slave-boson
representation
\cite{AC,Cole} of correlated
electrons in the dots.
In this representation, the creation (annihilation)
operator of electrons in the dot $m$ ($m=1,2$),
$d_{m\sigma}^\dag$($d_{m\sigma}$), is replaced by
$d_{m\sigma}^\dag\to f_{m\sigma}^\dag b_{m}^{}$
($d_{m\sigma}\to b_{m}^\dag f_{m\sigma}^{} $),
where $b_{m}$ ($f_{m\sigma}$) is the slave-boson
(pseudo-fermion) annihilation operator for an empty state
(singly occupied state).
We can thus model the system in
Fig. \ref{modelP} with a $N$(=2) fold degenerate Anderson Hamiltonian,
\begin{eqnarray}
H &=& \sum_{k_\alpha ,\sigma}\varepsilon _{k_\alpha\sigma}
c_{k_\alpha\sigma}^\dag c_{k_\alpha\sigma}^{}
 +\sum_{m,\sigma}\varepsilon _{m\sigma}
f_{m\sigma}^\dag f_{m\sigma}^{}
\nonumber\\
 &+& \frac{t_c}{N}\sum_{\sigma}(f_{1\sigma}^\dag b_1^{}
b_2^\dag f_{2\sigma}^{}+ h.c)
\nonumber\\
 &+& \frac{1}{\sqrt{N}}\sum_{k_\alpha,m,\sigma} (V_{\alpha m\sigma}
c_{k_\alpha\sigma}^\dag b_m^\dag f_{m\sigma}^{}+ h.c.)
\nonumber\\
 &+& \sum_{m}\lambda_m (\sum_{\sigma}f_{m\sigma}^\dag
f_{m\sigma}^{}+b_m^\dag b_m^{} -1).
\label{Hami}
\end{eqnarray}
The first and second terms in the Hamiltonian (\ref{Hami})
represent the electronic states in the leads and the dots,
where $c_{k_\alpha\sigma}^\dag$($c_{k_\alpha\sigma}$) is
the creation (annihilation) operator of an electron with
energy $\varepsilon _{k_\alpha\sigma}$ and spin $\sigma$
in the lead $\alpha$ ($\alpha=L, R$).
The coupling between the two dots (between the lead and the dot)
 is given by the third (fourth)
 term in the Hamiltonian (\ref{Hami}).
The last term with the Lagrange multiplier
$\lambda_m$ is introduced so as to
incorporate the constraint
imposed on the slave particles,
$\sum_{\sigma=\uparrow, \downarrow}
 {f_{m\sigma }^\dag f_{m\sigma }^{}}+b_m^\dag b_m^{}=1$.
The mixing term $V_{\alpha m\sigma}^{}$ in the 
Hamiltonian (\ref{Hami}) leads
to the linewidth function
\begin{equation}
\Gamma_{m,n\sigma}^\alpha(\varepsilon )=\pi \sum_{k_{\alpha}}^{}
V_{\alpha m\sigma}^{}V^*_{\alpha n\sigma}
\delta(\varepsilon -\varepsilon_{k_{\alpha}\sigma}).
\label{ganma}
\end{equation}
In the wide band limit, $\Gamma_{m,n\sigma}^\alpha(\varepsilon )$ 
is reduced to an energy-independent constant
 $\Gamma_{m,n\sigma}^\alpha$.

\begin{figure}[h]
\begin{center}
\includegraphics[scale=0.25]{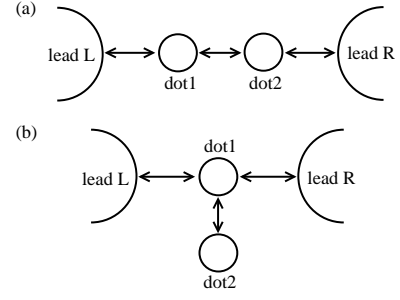}
\end{center}
\caption{
(a) Serial DQD system,
which is a special case of Fig. \ref{modelP}
($\Gamma_{1,1\sigma}^R=\Gamma_{2,2\sigma}^L=0$).
(b) T-shaped DQD system realized at 
$\Gamma_{2,2\sigma}^L=\Gamma_{2,2\sigma}^R=0$
in Fig. \ref{modelP}.
}
\label{modelST}
\end{figure}

According to Ref. \cite{Gueva}, we interpolate the serial DQD and
parallel DQD by continuously changing
$x=\Gamma_{1,1\sigma}^R=\Gamma_{2,2\sigma}^L$ 
with the resonance width
$\Gamma_{1,1\sigma}^L(=\Gamma_{2,2\sigma}^R)$ fixed as unity.
Note that at $x=0$ the model is reduced to  the DQD connected 
in series (serial DQD, Fig. \ref{modelST}(a)), and at $x=1$ the parallel DQD.
Similarly,  we modify the tunneling amplitudes in 
a different way, i.e. we change
$y=\Gamma_{2,2\sigma}^R=\Gamma_{2,2\sigma}^L$ by
keeping the resonance width
$\Gamma_{1,1\sigma}^L(=\Gamma_{1,1\sigma}^R)$ fixed as unity.
Then
 we can naturally interpolate the parallel  DQD ($y=1$) and
 the T-shaped DQD ($y=0$),
where only one of the two dots is connected to the leads,
as shown in Fig. \ref{modelST}(b).

To analyze the model, we apply the mean-field
approximation to the slave-boson treatment
\cite{AC,Cole}, 
in which boson fields are approximated by their static mean values,
 $b_m(t)/\sqrt{N}\to \langle b_m(t)\rangle /\sqrt{N}=\tilde{b}_m$.
We introduce the renormalized quantities
$\tilde{V}_{\alpha m\sigma}=V_{\alpha m\sigma}\tilde{b}_m$,
$\tilde{t}_c=t_c\tilde{b}_1\tilde{b}_2$,
and $\tilde{\varepsilon}_{m\sigma}=\varepsilon_{m\sigma}+\lambda_m$.
The mean-field values of
$\tilde{b}_1,\tilde{b}_2,\lambda_1$ and $\lambda_2$,
are determined by the following set of self-consistent equations, which are
derived by the equation of motion method
 for the non-equilibrium Keldysh  Green functions
\cite{Rosa,Lang},
\begin{eqnarray}
&&\tilde{b}_{1(2)}^{\,2}-i\sum_\sigma\int\frac{d\varepsilon }{4\pi}
G_{1,1(2,2)\sigma}^<(\varepsilon )=\frac{1}{2}
\label{self},
\\
&&(\tilde{\varepsilon}_{1(2)\sigma}-\varepsilon_{1(2)\sigma})
\,\tilde{b}_{1(2)}^{\,2}
\nonumber\\
&&\quad\quad
-i\sum_\sigma\int\frac{d\varepsilon }{4\pi}
(\varepsilon -\tilde{\varepsilon}_{1(2)\sigma})\,
G_{1,1(2,2)\sigma}^<(\varepsilon )=0.
\label{self2}
\end{eqnarray}
In the above equations, $G_{1,1(2,2)\sigma}^<(\varepsilon )$ is the Fourier
transform of the  Keldysh Green function
$G_{1,1(2,2)\sigma}^<(t-t')\equiv
i\langle f_{1(2)\sigma}^\dag (t)f_{1(2)\sigma}(t')\rangle$.
 Eq. (\ref{self}) represents the constraint imposed on
the slave particles, while Eq. (\ref{self2}) is obtained from
the stationary condition that the boson field is
 time-independent at the mean-field level. From the equation 
of motion  of the operator$f_{m\sigma}$
\cite{MW,Lang,Keldysh}, we
have the explicit form of the Green function,
\begin{widetext}
\begin{eqnarray}
G_{1,1(2,2)\sigma}^<(\varepsilon )=\frac{2i}{|D_{\sigma}(\varepsilon )|^2}
&\Big\{&\left(f_L(\varepsilon )\tilde{\Gamma}_{1,1(2,2)\sigma}^L
+f_R(\varepsilon )\tilde{\Gamma}_{1,1(2,2)\sigma}^R\right)
\left(\varepsilon -\tilde{\varepsilon}_{2(1)\sigma}
+i\tilde{\Gamma}_{2,2(1,1)\sigma}\right)
\left(\varepsilon -\tilde{\varepsilon}_{2(1)\sigma}
-i\tilde{\Gamma}_{2,2(1,1)\sigma}\right)
\nonumber\\
&+&\left(f_L(\varepsilon )\tilde{\Gamma}_{1,2(2,1)\sigma}^L
+f_R(\varepsilon )\tilde{\Gamma}_{1,2(2,1)\sigma}^R\right)
\left(\varepsilon -\tilde{\varepsilon}_{2(1)\sigma}
+i\tilde{\Gamma}_{2,2(1,1)\sigma}\right)
\left(\tilde{t}_c+i\tilde{\Gamma}_{2,1(1,2)\sigma}\right)
\nonumber\\
&+&\left(f_L(\varepsilon )\tilde{\Gamma}_{2,1(1,2)\sigma}^L
+f_R(\varepsilon )\tilde{\Gamma}_{2,1(1,2)\sigma}^R\right)
\left(\tilde{t}_c-i\tilde{\Gamma}_{1,2(2,1)\sigma}\right)
\left(\varepsilon -\tilde{\varepsilon}_{2(1)\sigma}
-i\tilde{\Gamma}_{2,2(1,1)\sigma}\right)
\nonumber\\
&+&
\left(f_L(\varepsilon )\tilde{\Gamma}_{2,2(1,1)\sigma}^L
+f_R(\varepsilon )\tilde{\Gamma}_{2,2(1,1)\sigma}^R\right)
\left(\tilde{t}_c-i\tilde{\Gamma}_{1,2(2,1)\sigma}\right)
\left(\tilde{t}_c+i\tilde{\Gamma}_{2,1(1,2)\sigma}\right)
\Big\},
\label{lessG}
\end{eqnarray}
\end{widetext}
with
$\tilde{\Gamma}_{m,n\sigma}^\alpha
=
\tilde{b}_{m}\tilde{b}_{n}\Gamma_{m,n\sigma}^\alpha $
and
$\tilde{\Gamma}_{m,n\sigma}
=
\tilde{\Gamma}_{m,n\sigma}^L+\tilde{\Gamma}_{m,n\sigma}^R
\,(m,n=1,2$ and $\alpha =L,R)$,
where the denominator of Eq. (\ref{lessG}) is
\begin{eqnarray}
D_{\sigma}(\varepsilon )
&=&
(\varepsilon-\tilde{\varepsilon}_{1\sigma}+i\tilde{\Gamma}_{1,1\sigma})
(\varepsilon-\tilde{\varepsilon}_{2\sigma}+i\tilde{\Gamma}_{2,2\sigma})
\nonumber\\
&&-(\tilde{t_c}-i\tilde{\Gamma}_{1,2\sigma})
   (\tilde{t_c}-i\tilde{\Gamma}_{2,1\sigma}).
\label{D}
\end{eqnarray}

 From the renormalized parameters determined self-consistently 
in Eqs. (\ref{self}) and (\ref{self2}),
we obtain the DOS
for the dot-1 and dot-2 as
\begin{eqnarray}
\rho_{1(2),\sigma} (\varepsilon )=-\frac{\tilde{b}_{1(2)}^{\, 2}}{\pi} 
\textrm{Im}
\left[
\frac{\varepsilon-\tilde{\varepsilon}_{2(1)\sigma}
+i\tilde{\Gamma}_{2,2(1,1)\sigma}}
{D_{\sigma}(\varepsilon )}
\right]_.
\label{DOS}
\end{eqnarray}
By applying the Landauer formula in steady state,
we can derive the current $I$ through the
two dots 
\cite{MW,Lang,Keldysh},
\begin{eqnarray}
I=\frac{2e}{h}\sum_{\sigma}
  \int d\varepsilon(f_L(\varepsilon )-f_R(\varepsilon ))
  T_{\sigma}(\varepsilon ),
\label{I}
\end{eqnarray}
where the transmission probability is given by
\begin{eqnarray}
T_{\sigma}(\varepsilon )
&=&
\frac{2}{D_{\sigma}(\varepsilon )}
\left[
\tilde{t}_c
 \left(\sqrt{\tilde{\Gamma}_{1,1\sigma}^L\tilde{\Gamma}_{2,2\sigma}^R}
 +\sqrt{\tilde{\Gamma}_{2,2\sigma}^L\tilde{\Gamma}_{1,1\sigma}^R}\right)
\right.
\nonumber\\
&&\!\!\!\!\!\!
\left.
+\sqrt{\tilde{\Gamma}_{1,1\sigma}^L\tilde{\Gamma}_{1,1\sigma}^R}
      (\varepsilon-\tilde{\varepsilon}_{2\sigma})
+\sqrt{\tilde{\Gamma}_{2,2\sigma}^L\tilde{\Gamma}_{2,2\sigma}^R}
      (\varepsilon-\tilde{\varepsilon}_{1\sigma})
      \right]_.^2
\nonumber\\      
\label{T}
\end{eqnarray}
We can thus compute the linear conductance $G_{V=0}=dI/dV|_{V=0}$ by
\begin{eqnarray}
G_{V=0}=\frac{2e^2}{h}\sum_{\sigma}
  T_{\sigma}(\varepsilon=0).
\label{IwithT}
\end{eqnarray}

\section{Numerical Results} \label{sec:result}

In this section, we discuss transport properties 
at zero temperature for the DQD systems with the  serial, parallel
and T-shaped geometries.
We also show the results obtained for the DQD systems  
connected to spin-polarized leads.  

\subsection{From serial to parallel DQD}

Let us first  discuss how the interference affects Kondo-assisted 
transport when we continuously 
change the system from the serial ($x=0$) to parallel
($x=1$) DQD. For simplicity, we deal with 
the symmetric dots in the Kondo regime
with $\varepsilon_{1}=\varepsilon_{2}=-3\Gamma_0$.
The band width of the leads is
taken as $D=60\Gamma_0$,
where $\Gamma_0=\Gamma_{1,1}^L=\Gamma_{2,2}^R$
(unit of energy), and the Fermi level of the leads
is chosen as the origin of energy.

\subsubsection{ non-polarized leads}

We begin with the DQD system
coupled to {\it non-polarized leads}: a brief report 
on this part can be found in Ref. \cite{tanaka2}.
We should also  mention two closely related works 
presented recently by Zhang et al. \cite{LuYu} 
and Dong et al. \cite{Lei}, who treated transport properties 
of similar models with non-polarized leads. 
Here, we briefly summarize 
our results, which will make clear how distinctly 
the interference appears in the parallel DQD 
with and without the spin polarization.

\begin{figure}[h]
\includegraphics[scale=0.35]{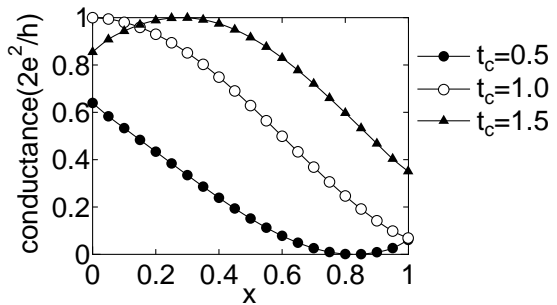}
\caption{
Linear conductance as a function of
$x=\Gamma_{1,1\sigma}^R=\Gamma_{2,2\sigma}^L$, where we 
set $\Gamma_{2,2\sigma}^R=\Gamma_{1,1\sigma}^L=1$.
}
\label{Ne3g}
\end{figure}

Figure \ref{Ne3g} shows the linear conductance
$G_{V=0}$ as a function of $x$
for several values of the inter-dot coupling $t_c$.
Starting from the serial DQD ($x=0$), we see that
 the conductance has a maximum
around $x=0.3$ for $t_c=1.5$, whereas it
  monotonically decreases for $t_c=1$ or
takes a tiny minimum structure around $x=0.8$ for $t_c=0.5$.
In any case, as the system approaches the parallel DQD 
($x \sim 1$), the conductance is considerably suppressed.
These characteristic properties 
 come from  the Kondo effect modified by the
interference, which is clearly seen in
the local DOS  and the transmission probability shown below.

\begin{figure}[h]
\includegraphics[scale=0.35]{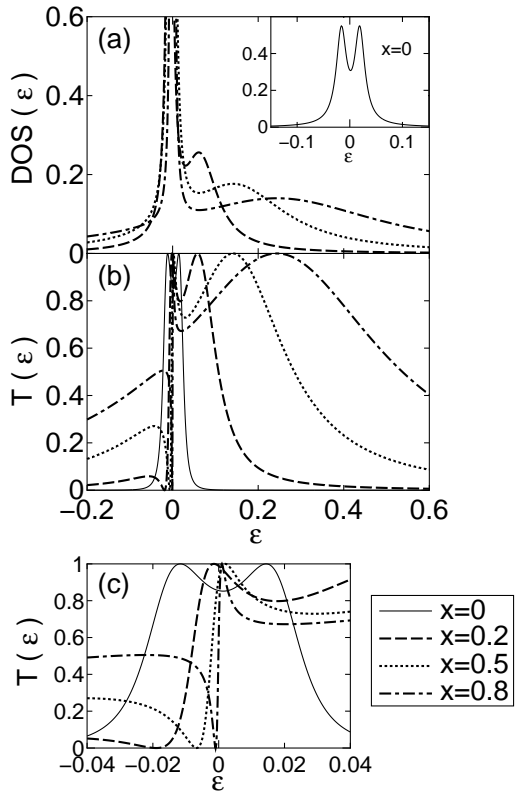}
\caption{
(a)DOS for $t_c=1.5$: $x=0.2, 0.5$ and $0.8$.
Inset shows the DOS at $x=0$.
(b)The transmission probability  $T(\varepsilon)$ for $t_c=1.5$.
(c)Enlarged picture of (b) around  $\varepsilon =0$.
}
\label{Ne3tc15}
\end{figure}
The local DOS of the dots 
around the Fermi energy $\varepsilon=0$ is shown
in Fig. \ref{Ne3tc15}(a) for a typical value of $t_c=1.5$.
For the serial case ($x=0$), the Kondo resonance has a
small splitting caused by the inter-dot
coupling $t_c$ \cite{Lang,George,Rosa,Eto,Eto2,Busser1,Izumi}.
The Kondo temperature in this case is roughly 
estimated as 0.02$\Gamma_0$, as seen in the inset 
of Fig. \ref{Ne3tc15}(a).
As  $x$ increases, one of the resonances
(lower-energy side), which is composed of 
 the "bonding" Kondo state of the DQD,
becomes sharp around the Fermi energy,
while the other "anti-bonding" Kondo state
(higher-energy side) is broadened above the Fermi level
\cite{Shimizu,LuYu,Lei,tanaka2,Busser2,Gueva}.
This tendency results from the fact that the bonding state
is almost decoupled from the leads while the anti-bonding
state is still tightly connected to the leads as $x \rightarrow 1$.
As a result, the renormalization of each Kondo temperature
occurs:
for instance, the Kondo temperature for the narrower (wider)
resonance is about $0.01\Gamma_0$ (0.2$\Gamma_0$) at $x=0.5$.

Such two modified Kondo resonances also appear in the 
transmission probability  $T(\varepsilon)$ (Fig. \ref{Ne3tc15}(b)).
A remarkable point is that 
 the sharp resonance in $T(\varepsilon)$ around the Fermi energy
in Fig. \ref{Ne3tc15}(b) becomes asymmetric and 
acquires a dip structure (more clearly seen 
in Fig. \ref{Ne3tc15}(c)).
The asymmetric peak, which may be regarded as a
Fano-like structure, is caused
by the interference between two conduction channels
having two distinct Kondo resonances. This type of interference
plays a crucial role to determine the Kondo-assisted conductance
in DQD systems \cite{Shimizu,LuYu,Lei,tanaka2,Busser2,Gueva}.
With the increase of $x$,
the maximum of the asymmetric resonance of $T(\varepsilon)$
in Fig. \ref{Ne3tc15}(c) passes through the Fermi energy,
from which we see why the linear conductance $G_{V=0}$
for $t_c=1.5$   in Fig. \ref{Ne3g} has a  maximum  
around $x=0.3$  and then decreases.
When we choose different values of $t_c$(=1.0, 0.5), analogous
interference effects occur, reducing the conductance substantially
when the system approaches the parallel DQD. Therefore, we see 
that apart from the detailed 
dependence, the characteristic behavior of the conductance in 
 Fig. \ref{Ne3g} is 
due to the Kondo-assisted transport modified by the interference 
effects.

Here we make a brief comment on the related work by Zhang et al.
\cite{LuYu}, who treated an analogous DQD system (two dots are connected 
via the exchange coupling $J$).  We have confirmed that 
for a given choice of 
the inter-dot coupling $t_c$ in the present model and 
$J$ in their model, the conductance exhibits similar interference 
effects. However, there are several different properties between
these two models. For example, in the serial-DQD  limit ($x=0$),
the former (latter) model has a finite (vanishing) conductance.
Also, the gate-voltage control of the dot levels 
shows an opposite tendency:
the increase of the dot-level enhances (suppresses) the 
effect of the inter-dot coupling $t_c$ ($J$). These different
properties  come from the fact 
that the former model allows charge fluctuations between the dots
while the latter model prohibits them.  We will come back to
this point later again in the discussions of the T-shaped DQD.

\subsubsection{spin-polarized leads}

We next discuss the influence of the {\it spin-polarized} leads
on the conductance, where the leads $L$ and $R$ have the same 
orientation of spin-polarization.
The spin-polarization of the leads gives rise to the difference
in the DOS between up- and down-spin conduction
electrons, so that the resonance width 
$\Gamma_{m,n\sigma}^\alpha$ should be spin-dependent.
To represent how large the
strength of the spin polarization is,
we introduce the effective
spin-polarization strength \textit{p}, following the
 definition given in the literature
\cite{Wang,Zhang,Sergu,Lopez,Koing,Utumi,Marti,JiMa,BinDon,
Choi,tanaka1,Julli,Slon}:
\begin{equation}
p=\frac{\Gamma_{m,m\sigma}^\alpha-\Gamma_{m,m\bar{\sigma}}^\alpha}
{\Gamma_{m,m\sigma}^\alpha+\Gamma_{m,m\bar{\sigma}}^\alpha}\,
 \,\, (0\le p\le 1)\,.
\label{p}
\end{equation}
This definition of \textit{p} is equivalent to 
the condition imposed on  $\Gamma_{m,m\sigma}^\alpha$,
\begin{eqnarray}
\Gamma_{m,m\uparrow}^{L(R)}&=&(1+p)\Gamma_{m,m}^{L(R)}|_{p=0},
\nonumber\\
\Gamma_{m,m\downarrow}^{L(R)}&=&(1-p)\Gamma_{m,m}^{L(R)}|_{p=0},
\label{resonance1}
\end{eqnarray}
for up-spin (majority-spin) electrons and
down-spin (minority-spin) electrons, respectively.
 We think that
the above simple assumption captures some essential 
effects due to the polarization of the leads.

Since the system for small $x$
shows the $p$-dependence similar to the serial case ($x=0$)
studied elsewhere \cite{tanaka1},
we discuss spin-polarized transport at $x=0.8$, which
includes essential properties inherent in
the parallel setup ($x\sim 1$).

\begin{figure}[h]
\includegraphics[scale=0.36]{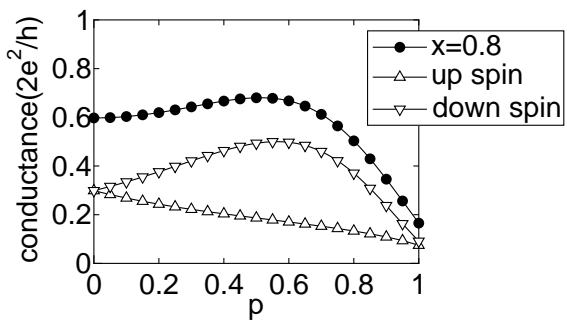}
\caption{Linear conductance as a function of
the spin-polarization strength $p$
for $t_c=1.5$ and $x=0.8$.
The contribution from up- and down-spin electrons is also shown.
}
\label{Pgtc15x08}
\end{figure}
\begin{figure}[h]
\includegraphics[scale=0.36]{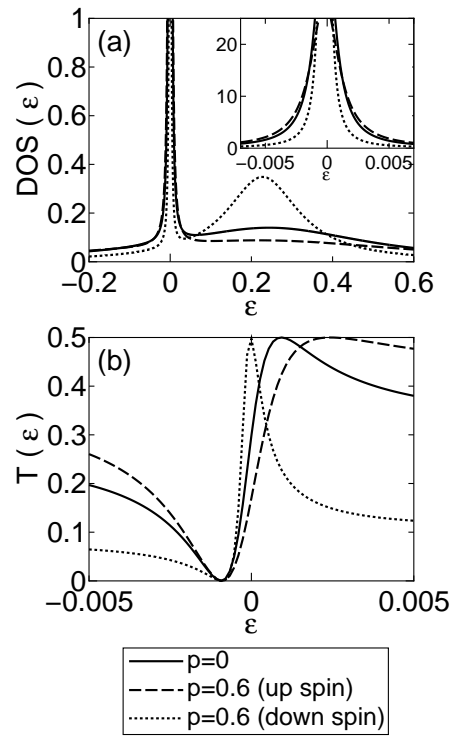}
\caption{
(a) DOS of up- and down-spin electrons connected to
normal leads ($p=0$)  and ferromagnetic leads
($p=0.6$).
We assume $t_c=1.5$ and $x=0.8$.
Inset: Enlarged picture around $\varepsilon =0$.
(b)Transmission probability $T(\varepsilon)$
of up- and down-spin electrons. 
}
\label{Ptc15x08}
\end{figure}

In Fig. \ref{Pgtc15x08} the linear conductance is shown 
as a function of the spin-polarization strength $p$
for the inter-dot coupling $t_c=1.5$ at $x=0.8$. 
As $p$ increases, the contribution of down-spin
electrons to the total conductance begins to increase,
 giving rise to a maximum around $p=0.6$, whereas
 the contribution of up-spin electrons
decreases  monotonically.
The above behavior due to the spin-polarization 
is explained in terms of the  modified 
Kondo resonances and  the asymmetric structure of $T(\varepsilon)$ 
around the Fermi energy. Shown in
Fig. \ref{Ptc15x08}(a) is the DOS of up- and
down-spin electrons for $p=0.6$, which is compared with
that for $p=0$.
The introduction of the spin-polarization of leads
broadens the Kondo resonance of
up-spin electrons whereas it sharpens that of 
down-spin electrons. Such change in the widths
of the Kondo resonances (inset of
Fig. \ref{Ptc15x08}(a)) modifies
the asymmetric nature of the transmission probability  $T(\varepsilon)$,
as seen in Fig. \ref{Ptc15x08}(b):
the asymmetric structure in $T(\varepsilon)$
of up-spin electrons for $p=0.6$ is smeared,
while that of down-spin electrons gets sharp.
The transmission probability at the Fermi energy for
up-spin electrons thus decreases with the increase of $p$, so that the
contribution of up-spin electrons to the
conductance goes down monotonically.
However, the conductance of
down-spin electrons takes a maximum  when
 the asymmetric peak of $T(\varepsilon)$ reaches 
the Fermi energy, as shown in Fig. \ref{Pgtc15x08}.
Similar considerations can be applied to
other choices of the parameters $t_c=0.5$ and $1.0$.
In any case in the Kondo regime, the spin polarization 
changes  the asymmetric structure of
$T(\varepsilon)$ more significantly for down-spin
(minority-spin) electrons,  making their contribution
to the conductance more dominant.

\subsection{From parallel to T-shaped DQD}
We now discuss how the Kondo-assisted transport changes its character
when the system is changed continuously
 from the parallel DQD to the  T-shaped DQD.
For this purpose, we change the resonance widths
$y=\Gamma_{2,2\sigma}^L=\Gamma_{2,2\sigma}^R$ by keeping
$\Gamma_{1,1\sigma}^L=\Gamma_{1,1\sigma}^R$ fixed as unity.
At $y=0$, the dot-2 is decoupled from two leads and
is connected only to the dot-1 via tunneling $t_c$ 
 (T-shaped geometry shown in Fig. \ref{modelST}(b)).
On the other hand, the system with $y\sim 1$ exhibits 
properties characteristic of the parallel DQD.
In the following calculation, we assume
 $\Gamma_{1,1\sigma}^L=\Gamma_{1,1\sigma}^R$
and take  $\Gamma_0=\Gamma_{1,1}^L|_{p=0}+\Gamma_{1,1}^R|_{p=0}$
as the unit of energy.

\subsubsection{ non-polarized leads}

\begin{figure}[h]
\begin{center}
\includegraphics[scale=0.33]{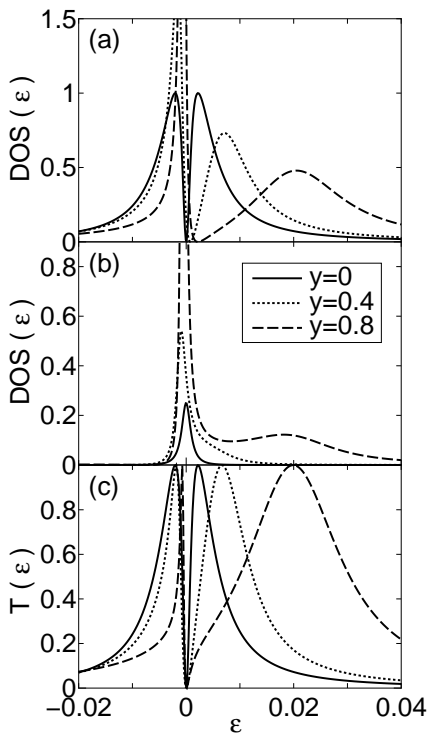}
\end{center}
\caption{
DOS of (a) dot-1 and (b) dot-2 around 
the Fermi energy.
(c)The corresponding transmission probability  $T(\varepsilon)$.
We set  $t_c=2$ and $\varepsilon_1 =\varepsilon_2 =-3.0$.
}
\label{PtoTdosTP}
\end{figure}

In Figs. \ref{PtoTdosTP}(a) and (b) the DOS 
 is shown  in the Kondo regime with
$\varepsilon_1 =\varepsilon_2 =-3.0$.
For $y=0.8$, the DOS has  the  Kondo resonance with the
double-peak structure both for the dot-1 and the dot-2, where
one of the two peaks is sharp around the Fermi energy
while the other is broad above the Fermi energy.
We have already encountered such situation in the previous
subsection for the system close to the parallel dot ($x \sim 1$).
However, when the system approaches the T-shaped DQD
($y\rightarrow 0$), quite different behavior emerges.
As $y$ decreases, the Kondo resonance of the dot-1 
gradually becomes symmetric
 with a sharp dip structure at $\varepsilon \sim 0$, whereas
 the DOS of the dot-2 develops a single Kondo peak located
at the same position as the dip structure in the DOS of the dot-1.
This change in the DOS can be interpreted as follows. 
 When $y \sim 1$ (close to the parallel
geometry) the two resonances are composed of the bonding and anti-bonding
Kondo states. On the other hand, for $y \sim 0$
(T-shaped geometry), they are composed of the Kondo states at the 
dot-1 and the dot-2, where the former (latter) has 
a broad (sharp) resonance
\cite{Tae,Kikoin,Taka,Busser2,Apel,Corna}.
As $y$ decreases, the double-peak structure of the Kondo resonances
gradually changes its properties between the two limits
 mentioned above. It should be 
noticed that in the T-shaped case, the DOS of the dot-1 itself develops
a dip structure by interference effects with the dot-2
\cite{Tae,Kikoin,Taka,Busser2,Apel,Corna}, 
 in contrast to the case of the  parallel geometry.

Figure \ref{PtoTdosTP}(c) shows the corresponding
transmission probability  $T(\varepsilon)$ for several
choices of $y$. As in the previous subsection, 
$T(\varepsilon)$ for $y=0.4$ and $0.8$ has
the asymmetric peak with the dip structure.
However, the dip  of $T(\varepsilon)$
 always stays around  $\varepsilon = 0$, whose position is 
determined by the Kondo state of the dot-2,
so that the conductance is almost zero (not shown here)
irrespective of the change of $y$.

So far, we have fixed the dot levels so as to keep the 
system in the Kondo regime.
We now discuss what happens when the energy levels of the dots 
are altered by the gate-voltage control.  To be specific,
we focus on the T-shaped DQD ($y=0$)
\cite{Tae,Kikoin,Taka,Busser2,Apel,Corna}. 
We first change $\varepsilon_1$  of the dot-1 by
keeping that of the dot-2 fixed as $\varepsilon_2 =-3.0$.
\begin{figure}[h]
\begin{center}
\includegraphics[scale=0.34]{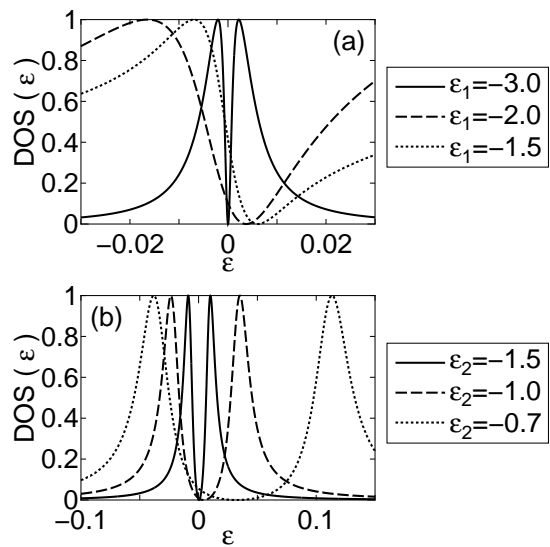}
\end{center}
\caption{
 DOS of the dot-1 for the T-shaped DQD ($y=0$): 
(a)  $\varepsilon_2=-3.0$ and
(b) $\varepsilon_1=-3.0$. The inter-dot coupling is 
chosen as  $t_c=2.0$.
}
\label{Tdose1}
\end{figure}

\begin{figure}[h]
\begin{center}
\includegraphics[scale=0.35]{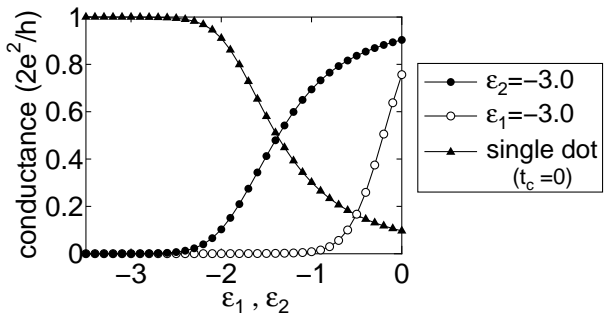}
\end{center}
\caption{
Linear conductance as a function of the bare energy level:
we change $\varepsilon_1$ (filled circles), by keeping
$\varepsilon_2=-3.0$ fixed, while  we change 
$\varepsilon_2$ (open circles)
by keeping  $\varepsilon_1=-3.0$  fixed.
The inter-dot coupling for
these plots is $t_c=2.0$.
For reference we plot the conductance 
for the single dot system ($t_c=0$)
as filled triangles.
}
\label{Tcone1}
\end{figure}

Figure \ref{Tdose1}(a) shows the DOS of the dot-1 plotted
as a function of $\varepsilon_1$.  The increase of $\varepsilon_1$
causes the broadening of the Kondo resonance
 and  enhances  charge fluctuations. Accordingly, 
the dip structure in  DOS of the dot-1 
shifts above the Fermi energy and gets more asymmetric. 
As a result, the conductance
gets large with the increase of $\varepsilon_1$,
as shown in Fig. \ref{Tcone1}.  This is  contrasted to
the single dot case, where the increase of $\varepsilon_1$
decreases the conductance. When the level  $\varepsilon_2$ of the 
dot-2 is altered, slightly different behavior appears.
In this case the increase of $\varepsilon_2$ does not
directly enhance charge fluctuations of the dot-1, but mainly
increases the renormalized tunneling $\tilde t_c$ because 
electron correlations between two dots get somewhat weaker.  This 
 merely enlarges the splitting of the double peaks in 
DOS of the dot-1 (Fig. \ref{Tdose1}(b)), and thus the conductance is
still very small, as seen  in the region
of $\varepsilon < -1$ in Fig. \ref{Tcone1}.  However,
as  $\varepsilon_2$ approaches the Fermi level, the enhanced
$\tilde t_c$ finally causes charge fluctuations of the dot-1,
and then increases the linear conductance.
In this way, the gate-voltage control of  $\varepsilon_1$
and  $\varepsilon_2$
appears in slightly different ways. Nevertheless, both exhibit
a similar tendency in the gate-voltage dependence of the conductance,
which is opposite to the  single-dot case. 

Although we have restricted our discussions
 to the T-shaped DQD here,
similar arguments about the gate-voltage control of the dot levels
can be straightforwardly applied to other cases such as the parallel DQD.

To conclude this subsection, we wish to mention 
some similarities and
differences between the present T-shaped DQD and 
a side-coupled QD that has been 
 studied theoretically
\cite{Liu,Kang,Torio,Aligia,Maru}
and experimentally
\cite{Sato}.
In both models, there exists the interference effect due 
to the T-shaped geometry including the side-coupled QD, which 
is essential to control transport properties. Namely, the 
coexistence of a direct tunneling without a side-dot and 
an indirect tunneling via a side-dot results in an asymmetric
transmission probability.
 In contrast to the T-shaped DQD, where
both of the two dots are highly correlated, 
the side-coupled QD is supposed to have
electron  correlations  only in the side dot.
In this sense, we would say that the charge-fluctuation regime with 
$\varepsilon \sim 0$ in the  T-shaped DQD, where electron 
correlations are somewhat suppressed in the dot-1, may exhibit 
properties similar to those in the side-coupled QD system.

\subsubsection{spin-polarized leads}

\begin{figure}[h]
\begin{center}
\includegraphics[scale=0.35]{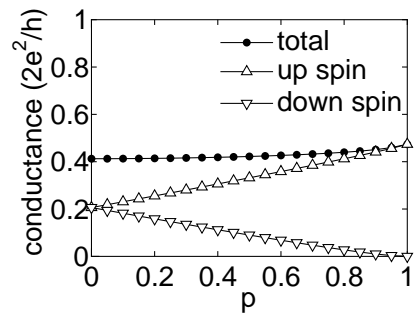}
\end{center}
\caption{ Linear 
conductance as a function of the
strength of the spin-polarization $p$. The contribution from
up- and down-spin electrons is also shown.
}
\label{TconP}
\end{figure}

We finally discuss the T-shaped DQD connected to the 
ferromagnetic leads, where the leads $L$ and $R$ have the 
same orientation of spin-polarization.
We will see below that the gate-voltage
control discussed above is helpful to understand the 
characteristic $p$-dependence of the conductance.
Here we consider the case of
$\varepsilon_1=-1.5$, $\varepsilon_2=-3.0$
and $t_c=2.0$, where the conductance is finite 
even at $p=0$, as shown in Fig \ref{Tcone1}.
\begin{figure}[h]
\begin{center}
\includegraphics[scale=0.36]{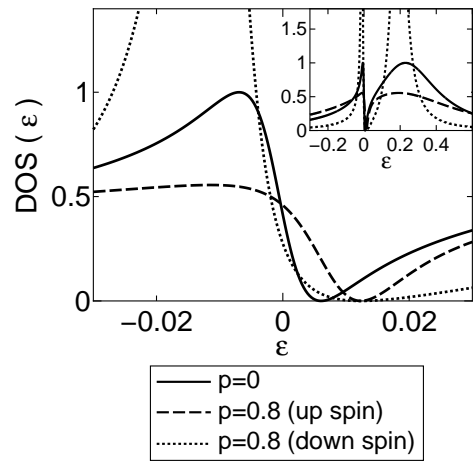}
\end{center}
\caption{
DOS of the dot-1 for the T-shaped DQD:
 non-polarized leads ($p=0$) and
spin-polarized leads ($p=0.8$).
We choose $\varepsilon_1=-1.5$, $\varepsilon_2=-3.0$ and
$t_c=2.0$.
Inset: DOS of the dot-1 drawn in a wider energy range.
}
\label{Tdos1P}
\end{figure}
Figure \ref{Tdos1P} shows the DOS of the dot-1 connected to
 non-polarized leads ($p=0$) and spin-polarized leads ($p=0.8$).
The width of the Kondo resonance (with a sharp dip)
is changed by the spin-polarization of the leads, so that the 
DOS of the dot-1 for up-spin (majority spin) electrons
is broadened while that for down-spin (minority spin) electrons
gets sharp. Accordingly, charge fluctuations 
of the up-spin (down-spin) electrons are somewhat enhanced 
(suppressed). Following the analysis done for the 
gate-voltage control (Fig.\ref{Tcone1}),
we naturally  expect that the contribution of up-spin
(down-spin) electrons to the total conductance
gets large (small) under the influence of the 
spin polarization.  We have indeed confirmed this tendency
in the conductance, where up-spin (down-spin) currents
are increased (decreased) as a function of $p$,
as shown in Fig. \ref{TconP}. 
Notice that this result
is contrasted to the $p$-dependence in Fig. \ref{Pgtc15x08}.
The difference, as mentioned above, comes from the fact
that the double structure of the Kondo resonances is
composed of the bonding and anti-bonding Kondo states
(the dot-1 and dot-2 Kondo states) in the case of 
Fig. \ref{Pgtc15x08} (Fig. \ref{TconP}), respectively.
To see the difference between Figs. \ref{Pgtc15x08} and \ref{TconP}
more clearly, we recall that in Fig. \ref{TconP} the increase of 
$p$ merely changes the resonance 
width and thus increases (decreases) $T(\varepsilon=0)$ for up-spin 
(down-spin) electrons, which results in the linear increase (decrease)
of the conductance as a function of $p$. On the other hand, as 
mentioned before, in the case of Fig. \ref{Pgtc15x08},
the introduction of $p$ not only changes the width of 
the asymmetric resonance in $T(\varepsilon)$ but also 
shifts its peak position for down-spin electrons, which leads to
the non-monotonic $p$-dependence of the conductance for down-spin
electrons. This consideration clarifies why we have encountered
the different $p$-dependence of the conductance in Figs. \ref{Pgtc15x08}
 and \ref{TconP}.

\section{Summary}

We have studied transport properties of the DQD systems 
with particular emphasis on the interplay of the Kondo
effect and the interference effect.  For this purpose we have 
observed how the Kondo-assisted conductance alters its 
properties
when the system is changed from the serial to parallel geometry, and
 from the parallel to T-shaped geometry.

For the serial DQD, it is known 
that the Kondo resonance may have a double-peak structure
due to the inter-dot tunneling, which somehow reduces the conductance.
 In this case, however, there is no
interference effect. On the other hand, when the system approaches 
the parallel DQD, there appear two distinct channels of electron
propagation via the bonding and anti-bonding 
dot states, which respectively form the sharp and broad Kondo 
resonances. The interference between
these two Kondo resonances gives rise to the asymmetric 
dip structure in the transmission probability, which
reduces the conductance significantly.
We have seen that the T-shaped DQD exhibits somewhat different 
properties. In this case, the DOS of the dot-1 connected to the leads
has a broader resonance and develops a sharp dip structure in 
it as a consequence of
interference with the dot-2 having  a much sharper Kondo 
resonance.  Therefore,  the DOS as well as
the transmission probability always have the dip structure
at the Fermi level, thus giving a very small conductance
 in the Kondo regime.

It has been shown that the gate-voltage control 
causes two main effects via the change of the dot-levels:
the increase of the dot-level induces charge fluctuations and
also causes 
the renormalization of the effective inter-dot coupling. These
effects make the characteristic behavior of the 
conductance quite different from that for the
single dot case.  In particular, it has been demonstrated 
in the T-shaped case that the gate-voltage (i.e. dot-level) control
leads to a tendency contrary to the single-dot case: 
the conductance increases with the increase of 
the bare level of the dot, reflecting
the Kondo effect influenced by the interference effect.

We have also discussed the  
impact of the FM leads on transport properties.
The change in the Kondo resonances due to 
the spin-polarization of the FM leads 
modifies the asymmetric structure in the transmission 
probability, so that the spin-dependent currents may
flow depending on the spin-polarization strength $p$.
It has been shown that the spin-dependent conductance is
quite sensitive to the geometry of DQD, implying that
 the interference of electrons plays a crucial role to
determine the transport properties.

 Experimentally, the Kondo effect in DQD systems
has been observed recently
\cite{Jeong,Chen}.
It is thus expected that interference effects
 will be
systematically studied in a variety of 
DQD systems including the T-shaped DQD. 
We also hope that the spin-dependent conductance in such correlated
DQD can be controlled by
 spin-polarization in the near future.

\begin{acknowledgments}
The work is partly supported by a Grant-in-Aid from the
 Ministry of Education, Culture,
Sports, Science and Technology of Japan.
\end{acknowledgments}

\end{document}